# Response of AGATA Segmented HPGe Detectors to Gamma Rays up to 15.1 MeV

F.C.L. Crespi[a,b], R. Avigo[a,b], F. Camera[a,b], S. Akkoyun[c], A. Ataç[c], D. Bazzacco[d], M. Bellato[d], G. Benzoni[b], N. Blasi[b], D. Bortolato[d,e], S. Bottoni[a,b], A. Bracco[a,b], S. Brambilla[b], B. Bruyneel[f], S. Ceruti[a,b], M. Ciemała[g], S. Coelli[b], J. Eberth[h], C. Fanin[i], E. Farnea[d], A. Gadea[j], A. Giaz[a,b], A. Gottardo[i], H. Hess[h], M. Kmiecik[g], S. Leoni[a,b], A. Maj[g], D. Mengoni[d,k], C. Michelagnoli[d,e], B. Million[b], D. Montanari[d,e], L. Pellegri[a,b], F. Recchia[d,e], P. Reiter[h], S. Riboldi[a,b], C.A. Ur[d,e], V. Vandone[a,b], J.J. Valiente-Doboni[i], O. Wieland[b], A. Wiens[h]

and The AGATA Collaboration

[a] *Dipartimento di Fisica, Università di Milano, I-20133 Milano, Italy*
[b] *INFN, Sezione di Milano, I-20133 Milano, Italy*
[c] *Department of Physics, Faculty of Science, Ankara University, 06100 Tandoan, Ankara, Turkey*
[d] *INFN, Sezione di Padova, I-35122 Padova, Italy*
[e] *Dipartimento di Fisica dell'Università, Sezione di Padova, I-35122 Padova, Italy*
[f] *CEA-Saclay DSM/IRFU/SPhN, 91191 Gif-sur-Yvette, France*
[g] *The Niewodniczanski Institute of Nuclear Physics, Polish Academy of Sciences, 31-342 Krakow, Poland*
[h] *Institut für Kernphysik, Uni zu Köln, Zülpicher Str. 77, D-50937 Köln, Germany*
[i] *INFN, Laboratori Nazionali di Legnaro, IT-35020 Legnaro, Italy*
[j] *IFIC, CSIC-University of Valencia, ES-46071 Valencia, Spain*
[k] *School of Engineering, University of the West of Scotland, Paisley PA1 2BE, United Kingdom*

**Abstract**

The response of AGATA segmented HPGe detectors to gamma rays in the energy range 2-15 MeV was measured. The 15.1 MeV gamma rays were produced using the reaction d($^{11}$B,n$\gamma$)$^{12}$C at $E_{beam}$ = 19.1 MeV, while gamma-rays between 2 to 9 MeV were produced using an Am-Be-Fe radioactive source. The energy resolution and linearity were studied and the energy-to-pulse-height conversion resulted to be linear within 0.05%. Experimental interaction multiplicity distributions are discussed and compared with the results of Geant4 simulations. It is shown that the application of gamma-ray tracking allows a suppression of background radiation following neutron capture by Ge nuclei. Finally the Doppler correction for the 15.1 MeV gamma line, performed using the position information extracted with Pulse-shape Analysis, is discussed.

## 1. Introduction

In many in-beam gamma spectroscopy experiments the detection of high-energy gamma-rays in the 10-20 MeV range is of primary importance (see e.g. [1-5]). The limited size of the presently available HPGe crystals (up to ~400 cm$^3$) affects the possibility to detect the full energy deposition of such high-energy photons. However, large detection volumes (and consequently large detection efficiencies) can be obtained by using composite germanium detectors, namely using multiple crystals within the same cryostat as was done in the past with the Clover detectors [6] and with the EUROBALL Cluster detectors [7-10]. The response function of these latter detectors was investigated up to 15 MeV [11-13]. The added benefit of generating large detection volume through several small crystals is the reduction of the Doppler broadening of lines induced by the finite solid angle subtended by each crystal in case the photons are emitted from recoiling nuclei. With the new generation high-resolution gamma-ray spectrometers like AGATA [14-16] and GRETA [17,18], the HPGe crystals are operated in position-sensitive mode through a combination of electrical segmentation of the outer electrodes, digital electronics and sophisticated Pulse Shape Algorithms [19-28]. The energy and direction of the individual photons are extracted through dedicated gamma-ray tracking algorithms [29-32]. It should be remarked that the individual interaction points are extracted with sub-segment precision, which experimentally turns out to be better than a 3D Gaussian with 5 mm FWHM in each direction (see for instance [33-36]). In order to achieve this goal, remarkable effort has been concentrated on the characterization of highly-segmented HPGe detectors [37-52].

The possibility to improve the performances of a gamma-ray spectrometer at high energies using accurate 3D position information was first proposed in ref [53].

The performance of the Advanced GAmma-ray Tracking Array (AGATA) detectors with in-beam tests were discussed in ref [33-36]. These studies, however, were limited to gamma-rays up to 4 MeV. The present work provides the first detailed study of the response of AGATA detectors to gamma-rays up to 15.1 MeV. This study represents an important test of the AGATA detectors for the measurement of high-energy gamma-rays, in terms of energy resolution, tracking efficiency and performance of the PSA algorithms. This aspect will be important in the forthcoming experimental campaign with relativistic beams [54] at GSI. In fact, in this case, the energies of the gamma-rays emitted in flight can be significantly Doppler shifted towards higher values.

In section 2 we describe the experimental set up, the Am-Be-Fe source calibrations and the in-beam test. In section 3 the results concerning detector energy resolution and linearity as a function of the gamma-ray energy are presented. Experimentally extracted interaction multiplicity distributions are shown and compared with Geant4 [55-57] simulations in section 4. Finally, in section 5 we discuss the Doppler correction using the PSA and gamma-ray tracking for the 15.1 MeV gamma line.

## 2. In-beam test and Am-Be-Fe source measurement

The reaction used to produce the 15.1 MeV gamma-ray was

$$d(^{11}B,n\gamma)^{12}C \text{ at } E_{beam} = 19.1 \text{ MeV}.$$

A $^{11}$B beam with an energy of 45 MeV from the Legnaro XTU Tandem accelerator… was degraded to 19.1 MeV using a golden foil in front of the target (29 mg/cm$^2$). The reaction populates the resonance state at 15.1 MeV in $^{12}$C nucleus which is produced with a v/c of ~5%. This state decays directly to the ground state (with a branching ratio of 92% [58]) by emitting a single M1 gamma ray with an energy of 15.1 MeV [59-61]. The target was made of $C_{32}D_{66}$ (dotriacontane-d66) material with a thickness of 490 μg/cm$^2$ deposited on a 0.1 mm thick tantalum backing. Both the recoiling nuclei and the beam were stopped in the target backing. The gamma rays produced in the reaction were measured with two AGATA triple clusters, which were placed at a distance of 13.5 cm from the target. The AGATA electronics was set in order to have 0-20 MeV dynamic range. The trigger condition did not require any coincidence with other detectors. One large volume cylindrical 3.5" x 8" LaBr$_3$:Ce detector, having larger efficiency as compared to one single Agata crystal and operated using an independent acquisition system, was added to the experimental set-up for monitoring purposes (upper panel of Figure 1) [62,63]. Before the in-beam measurement the detectors were calibrated using the Am-Be-Fe source. The Am-Be-Fe source was placed into a 3 x 3 cm hole drilled in an iron slab of dimension 7 x 7 x 20 cm and surrounded by paraffin wax in a cylindrical shape (20 x 20 cm), see the bottom panel of Figure1. The neutrons from the Am-Be-Fe source were thermalized in the paraffin housing and then captured in iron producing gamma-rays up to 9.3 MeV.

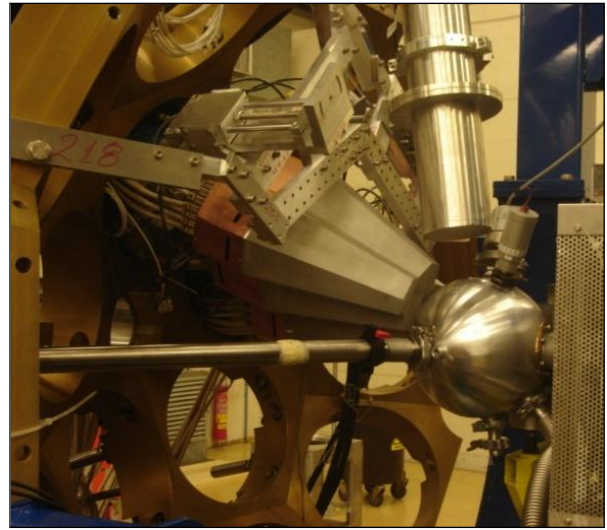

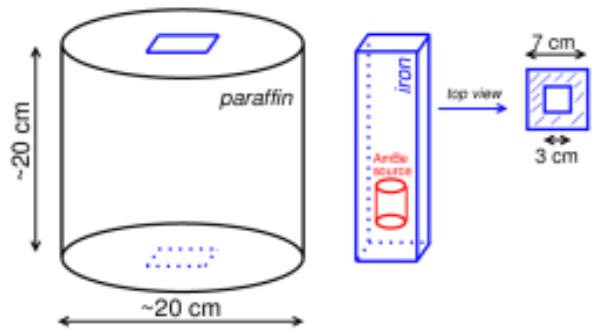

Fig. 1. Upper panel: The experimental set-up consisting of two AGATA triple clusters and one 3.5"x8" cylindrical LaBr$_3$:Ce scintillation detector. Lower panel: schematic representation of the Am-Be-Fe source.

The gamma-ray spectrum acquired using the Am-Be-Fe source is displayed in Figure 2 and the gamma lines used for the analysis are labeled according to the reaction which originated them. These data allowed the calibration of the detectors and a check of the linearity and energy resolution of the AGATA detectors. The average counting rate per crystal was 0.9 kHz for the case of source measurement and 1.2 kHz for the in-beam test.

## 3. Energy resolution and linearity

In Figure 3 the relative energy resolution (i.e. FWHM /$E_{gamma}$) as a function of the gamma-ray energy is displayed. The data associated to the single crystal showing the best performance are reported with empty black circles. The black triangles represent, instead, the energy resolution obtained with a sum of the energies detected by the crystals that fired in the event (add-back).

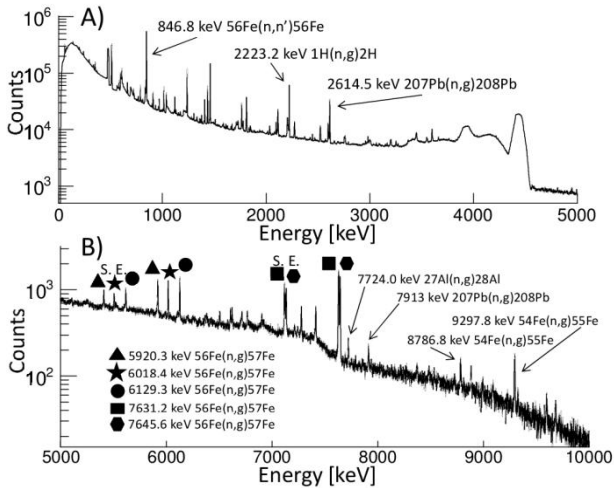

Fig. 2. Gamma-ray energy spectrum measured with an Am-Be-Fe source in the 0-5 MeV range (panel A) and in the 5-10 MeV range (panel B). The gamma lines used for the analysis are labeled indicating the reaction that originated them.

All the spectra analyzed in this section were extracted without using any kind of filter, just summing up the energy measured in each segment separately; this procedure is feasible because of the low gamma-ray multiplicity (see e.g. section 4). These segment energies are extracted at pre-processing level by applying the moving window deconvolution (MWD) algorithm [64,65] on the incoming data streams. In this way it was then possible to perform (offline) a fine gain matching for all segments. This latter procedure resulted to be extremely important especially when high energy gamma rays are involved. In addition, for each crystal, the sum energy of the segments was forced to be equal to the energy extracted from the core signal. This was done to recover the segment energy resolution, degraded due to neutron damage [66]. It has to be mentioned that a more sophisticated method to recover neutron damage in segmented HPGe detectors, exploiting position information provided by PSA algorithms, was recently developed [67]. However we don't expect this method to provide significant improvement for the specific case of the high energy gamma-rays considered in this work.

As can be seen from Figure 3 the experimental data follow the expected $E^{-1/2}$ trend (indicated by the black dashed line). The FWHM of the highest energy gamma line (i.e. 9297.8 keV) is 6.1 keV for the case of single crystal showing the best performance and 7.6 keV for the add-back. The energy resolution obtained for the 15.1 MeV gamma emitted in the in-beam test is not displayed since the FWHM of the peak is, in this case, dominated by the Doppler broadening induced by the reaction mechanism (see section 5 for details). However, considering the trend showed by the data displayed in figure 3, an intrinsic resolution of the order of 10 keV should be expected at the energy of 15 MeV.

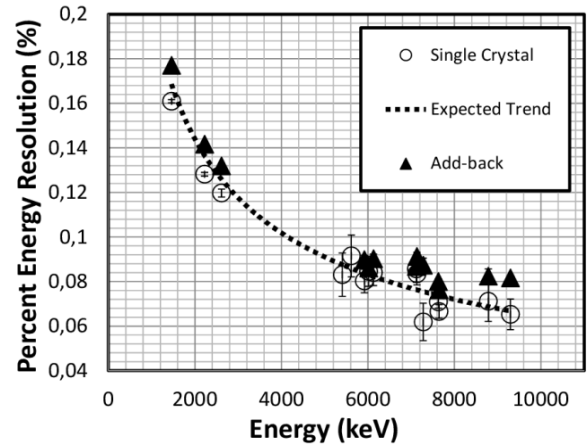

Fig. 3. The relative energy resolution of the AGATA detector is given for the Am-Be-Fe source data. The data for the single detector showing the best performances are reported in empty black circles. The black triangles represent instead the energy resolution for the add-back performed among all the crystals that fired in each event. The experimental data follow the expected $E^{-1/2}$ trend (indicated by the dashed black line).

In the following we present the study of the linearity for the energy to pulse height conversion up to 15 MeV.

The plot in Figure 4 displays the measured energy versus the tabulated energy for gamma lines of the Am-Be-Fe source and for the 4.4 MeV and 15.1 MeV gamma rays from the in-beam measurement. The measured energy is obtained with a linear calibration using the 1172 keV and 1332 keV lines of $^{60}$Co source. Correction factors for segment energies gain matching were extracted then using linear interpolation of the 846.8 keV, 2223.2 keV and 2614.5 keV gamma lines in the spectra in which the most energetic release took place in the selected segment.

In the plot displayed in Figure 5 the deviation from perfect linearity is displayed as a function of the energy. This is determined as the ratio between the difference of the measured and tabulated energy with the measured energy (Deviation = $(E_{meas}-E_{tab})/E_{meas}$). The data with the larger error bars are associated to the gamma rays emitted in flight (acquired during the in-beam test). The total deviations from ideal linearity are lower than 0.1 % in the energy range 2 – 15 MeV. The results are consistent with those reported in [13] for the case of EUROBALL [7-10] clusters.

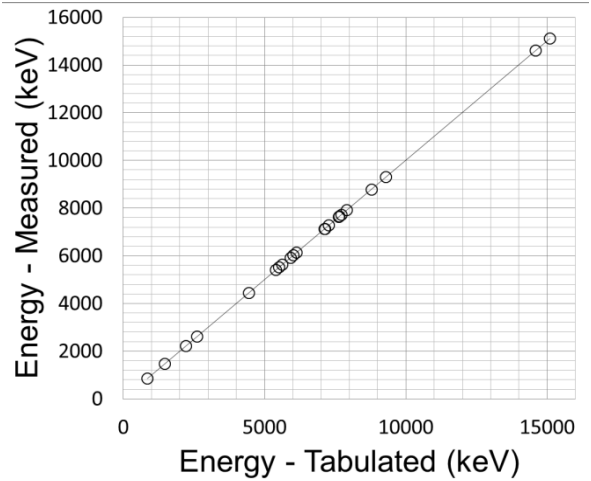

Fig. 4. Energy tabulated versus the measured energy for gamma lines of the Am-Be-Fe source and for the 4.4 MeV and 15.1 MeV gammas from the in-beam test as well.

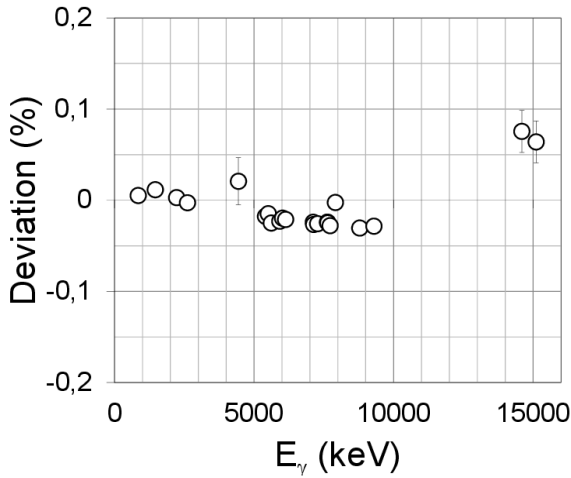

Fig. 5. Deviation of the measured energies from the energy tabulated for each gamma line of the Am-Be-Fe source and for the 4.4 MeV and 15.1 MeV gammas from the in-beam test. If not displayed error bars are smaller than symbol size.

## 4. Multiplicity distributions

In this section the multiplicity distributions of AGATA clusters, crystals and segments are discussed. The results presented in this section were extracted using data from the Am-Be-Fe source measurement described in section 2. Unless otherwise specified, the plots are produced without applying any filter to the data (e.g. gamma-ray tracking algorithm).

In table 1 the cluster multiplicity distributions for full energy peak (FEP) and background events are listed. The table clearly shows a general increase with gamma-ray energy of the fraction of the events in which the energy release is shared between both clusters ($M_{clust}=2$). Background events show a larger percentage of $M_{clust}=2$ events as compared to full energy peak ones. The same behavior can be observed in Figure 6, which displays the crystal multiplicity distributions for full energy peak (bottom panel) and background events (top panel). Such a behavior, in the case of the Am-Be-Fe source data, is due to the fact that background events originate mostly from neutron interactions in HPGe detectors and subsequent neutron induced gamma emission. These events are expected to have an average larger multiplicity as compared to gamma-ray FEP events leading to the same total energy release in the HPGe detectors. This can be attributed to the presence of additional interaction points associated to inelastic neutron scattering with Ge nuclei [68] and to the multiplicity of gamma-rays emitted following Ge nuclei de-excitation.

Figure 7 displays the centroid of segment multiplicity distributions as a function of gamma-ray energy for FEP (top panel) and background (bottom panel) events. In addition, the segment multiplicity distributions extracted using a simple add-back algorithm (i.e. summing up the energies of all the interactions in the 2 clusters) are compared with those extracted applying the gamma-ray tracking alogirthm [69]. It should be mentioned here that AGATA detectors can provide also sub-segment information concerning interaction number distributions (see e.g [25]). Nevertheless in this specific study the used algorithm [19] provides a single interaction point per segment where a net charge deposition took place, implying that the multiplicity distributions of interaction points and of segments necessarily coincide.

By looking at Figure 6 it can be noted that even though the general behaviour is identical up to 7 MeV, for higher energies a clear deviation between the two curves appears. This effect can be attributed to the background suppression performed by the tracking algorithm, rejecting neutron capture events characterized by events of high multiplicity emitted following Ge nuclei de-excitation. This effect can be directly observed in the suppression, performed by the tracking algorithm, of the 10.196 MeV line shown in Figure 8.

| Full Energy Peak (FEP) Events | | |
|---|---|---|
| Energy (MeV) | $M_{clust} = 1$ | $M_{clust} = 2$ |
| 2.2 | 92% | 8% |
| 4.4 | 88% | 12% |
| 7.6 | 85% | 15% |
| 9.3 | 86% | 14% |

| Background | | |
|---|---|---|
| Energy (MeV) | $M_{clust} = 1$ | $M_{clust} = 2$ |
| 2.2 | 86% | 14% |
| 4.4 | 80% | 20% |
| 7.6 | 65% | 35% |
| 9.3 | 58% | 42% |

Table 1. Cluster multiplicity for FEP and background events. Two AGATA triple clusters were used in the measurement.

The spectra were obtained by applying gamma-ray tracking algorithm (red line spectrum) and the add-back one (black line spectrum). The peak that appears in the add-back spectrum is associated to the sum energy of the gamma-rays emitted following the $^{74}$Ge nucleus de-excitation, after neutron capture by $^{73}$Ge. The ground state decay from 10.196 MeV level is not allowed [70,71], therefore the events in the peak have gamma multiplicities larger than one. As the tracking algorithm [69] recognizes the peak as a sum-peak of two or more gamma-rays it is suppressed in the 'Tracking' spectrum.

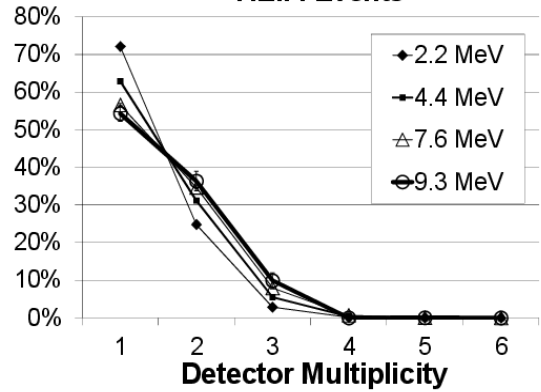

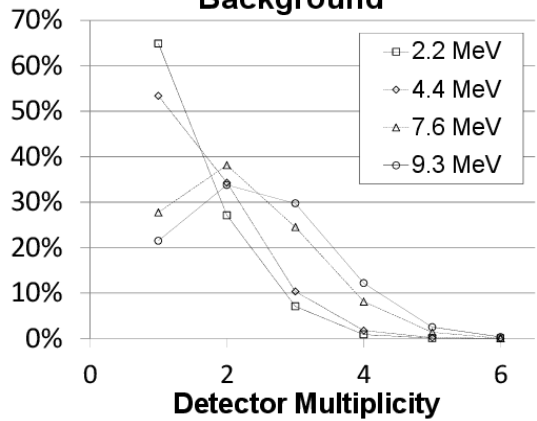

Figure 6. Crystals Multiplicity for FEP and background events. If not displayed error bars are smaller than symbol size.

In figure 9 segment multiplicity distributions for the cases of full energy peak ($E_{gamma}$ = 7.6 MeV), single escape and double escape events are compared. The fact that the distributions have centroids shifted toward higher multiplicities for the case of full energy and single escape is due to the presence of the 511 keV gamma-rays from pair production,. The fact that ~50% of double escape events have multiplicity larger than one can be adduced to the presence of bremsstrahlung radiation and Compton interactions of the gamma-ray prior to the pair production. In Figure 9 the results of Geant4 simulations [56,57] are also reported, showing a good matching with the experimental distributions.

## 5. Doppler correction of 15.1 MeV gamma-rays

In contrast to the Am-Be-Fe radioactive source data, the 15.1 MeV gamma-rays are emitted by a $^{12}$C nucleus moving at v/c ~ 5% (see section 2) and thus the energy of the gamma-rays detected in the laboratory system is shifted according to the expression:

$$E_{\gamma,Shifted} = E_{\gamma 0} \frac{(1-\beta^2)^{1/2}}{(1-\beta \cos\theta)} \quad (1)$$

where: $E_{\gamma 0}$ is the energy of the gamma-ray in the rest frame of the nucleus, β is the velocity of the nucleus in the laboratory system relative to the speed of light and θ is the angle between the direction of motion of the nucleus and the emission direction of the gamma-ray. While the angular distribution of the $^{12}$C recoils is not measured by our detection system, with the AGATA detectors it is possible to determine the emission direction of the detected photon at different levels of precision, namely: i) using the central position of the crystal with the largest energy deposit, ii) the central position of the segment with the largest energy deposit, iii) the position of the most energetic interaction point provided by the PSA algorithm [19] (from now on we refer to this procedure as "PSA+1HitID"), iv) the incoming direction provided by the gamma-ray tracking algorithm [69].

The PSA+1HitID algorithm calculates, for each event, the sum energy in all the detectors and determines the direction of the detected gamma ray starting from the assumption that the first interaction corresponds to the location of the most energetic interaction [75] extracted by PSA algorithm [19].

This solution was chosen since the efficiency of the standard tracking algorithm [69] was found to significantly decrease in the 10-20 MeV energy range. In particular after applying the mgt [69] tracking algorithm on both simulated and experimental data it resulted that the ratio between the events in the 15.1 MeV full energy peak for the tracked spectrum and the standard add-back with PSA+1HitID is 0.25. This is related to the fact that the used tracking algorithm was not optimized to treat gamma rays in the 10-20 MeV range where the pair production becomes the dominant interaction mechanism. In addition, in the present in-beam test the 15.1 MeV gamma-ray is produced by the direct decay into the ground state of $^{12}$C, therefore the multiplicity is always one. This fact allows a simpler approach as the PSA+1HitID to give the best results.

It is important to stress that the "multiplicity = 1" condition is fulfilled in several AGATA physics cases where the measurement of high-energy gamma rays is required (e.g. the measurement of the Pygmy Dipole Resonance [1]).

In the used reaction (see section 2) $^{12}$C is produced with a β of ~5%, however the velocity of the $^{12}$C ions was not measured. Therefore in order to Doppler correct in the optimal way the detected gamma-ray energy we determined the value of β which better optimizes the centroid and width of the 15.1 MeV full energy peak. In such a way we extracted an averaged velocity vector of magnitude 0.046 (β) and components (0, 0.85, 0.51) in the AGATA frame of reference; the AGATA reference frame is a right handed reference frame where the z axis coincides with the optical axis of PRISMA and x axis points downward (see [15,36,56,57]).

The components of the velocity vector are compatible with the beam direction. It is interesting to note that the best value of the extracted velocity is consistent with the results of simulations of the $^{12}$C ion velocity distribution performed with PACE4 [72-74] giving a mean β of 0.048. More specifically we found that the 95% confidence interval for the β value is between 0.042 and 0.058 and between 0° and 10° for the deviation angle with respect to the beam direction eam direction in the AGATA frame of reference.

The spectra in the region of 15 MeV are shown in the panels of figure 10. In particular different Doppler corrections were applied, using as gamma-ray emission direction the different options listed at the beginning of this section. In the top panel of Figure 10 the spectrum obtained without Doppler correction (dashed black line) compared to: i) the spectrum obtained by applying the Doppler correction using the central position of the HPGe crystal with the largest energy deposit (thick gray line), ii) the spectrum obtained by applying the Doppler correction using the central position of the segment with the largest energy deposit (thin blue line) and iii) the spectrum obtained by using the full information provided by the PSA "PSA+1HitID" (thin red line). By looking at the spectra displayed in the bottom panel of figure 10 one can note the marked improvement in the FWHM of the 15.1 MeV peak passing from the spectrum using only the central position of each crystal (> 160 keV FWHM), (i.e. detectors operated in standard mode) to the "PSA+1HitID" (red line, 119 keV FWHM) (see also table 1).

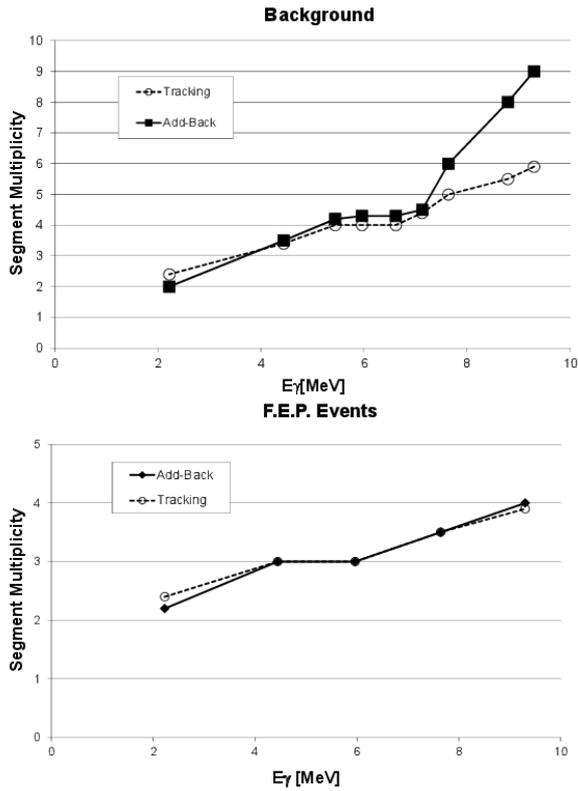

Figure 7. Segments multiplicity as a function of gamma energy for FEP and background events. Results for gamma-ray tracking and standard "add-back" are compared. The centroids of each distribution are plotted with empty circles and black squares respectively. Error bars are smaller than symbol size.

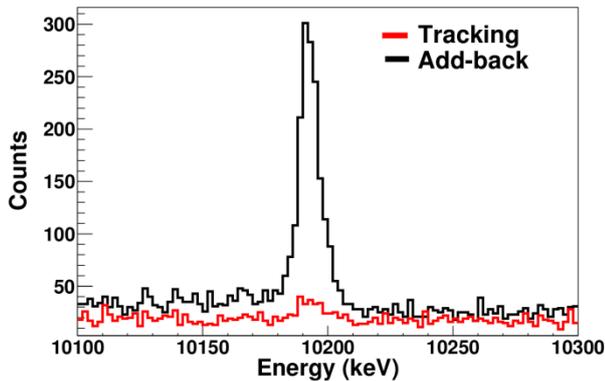

Figure 8. The spectra obtained using the gamma-ray tracking algorithm (red line) and the standard add-back one (black line) are displayed in the region around 10.196 MeV (i.e. $^{74}$Ge neutron separation energy). The peak that appears in the add-back spectrum is associated to the sum energy of the gamma rays emitted following the $^{74}$Ge nucleus de-excitation, after neutron capture by $^{73}$Ge. These events are correctly recognized as composed by multiple gamma rays and disentangled by the tracking algorithm.

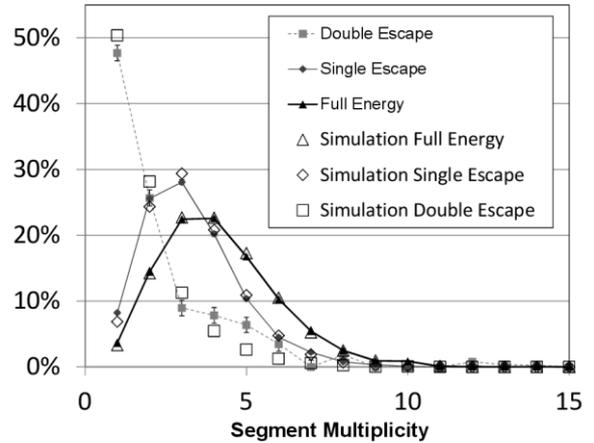

Figure 9. Segments multiplicity distributions for 7.6 MeV gamma-rays. The case of FEP, SE, DE are compared. The resutls of the corresponding Geant4 simulations (add-back) are shown. If not displayed error bars are smaller than symbol size.

It is important to stress that, in this particular case, PSA techniques do not improve in a significant way the energy resolution as compared with the spectrum where Doppler correction was made using segment centers. In fact the FWHM slightly improves from 122 keV to 119 keV (see table 2). This fact is due to the uncertainty in $^{12}$C ion vector velocity. The missing reconstruction on event-by-event basis of the $^{12}$C ion velocity vector represents in this case the main limiting factor in the Doppler broadening correction quality.

In order to verify the different contributions to the final width (119 keV) of the 15.1 MeV peak Geant4 simulation were performed and compared to the experimental result, see Figure 11 This simulation was performed using the AGATA code [56,57], applying then the same algorithm used to process the experimental data. The $^{12}$C ion velocity distribution was calculated using PACE4 [72-74] as discussed earlier. In the simulation the value of the intrinsic energy resolution of the detectors was extrapolated using the $E^{-1/2}$ law (see Figure 3) and set to 8 keV at 15.1 MeV. It should be pointed out, however, that this value has negligible impact on the final energy resolution obtained in the experimental spectrum (see Table 2), since this is dominated by the Doppler broadening effect.

As can be noted by looking at Figure 11 there is good agreement between the two curves, confirming that the measured FWHM of the Doppler corrected 15.1 MeV gamma line to 119 keV is understood.

| FWHM of 15.1 MeV peak | |
|---|---|
| PSA+1HitID | 119 keV |
| Segments | 122 keV |
| Crystals | >160 keV |

Table 2. Values for the FWHM of the 15.1 MeV gamma line obtained with Doppler correction using different position information, as described in the text. The main factor limiting the FWHM of the 15.1 MeV gamma line was found to be the uncertainty due to the missing event by event reconstruction of the $^{12}$C ion velocity vector. However, it is important to point out that considering the trend showed by the data displayed in figure 3, an intrinsic resolution of the order of 10 keV should be expected at the energy of 15 MeV.

## 6. Conclusions

In this paper we studied the response of two AGATA triple clusters to gamma-rays in the energy range 2-15 MeV. The energy resolution was found to scale as $1/\sqrt{E}$, once an accurate gain matching of the segments is performed. The linearity resulted to be better than 0.05% up to 10 MeV and better than 0.1% up to 15.1 MeV. The experimental interaction multiplicity distributions show that, for high energy gamma-rays, background events are characterized by higher average multiplicity than full energy peak ones. This is related to neutron capture events which characterize the spectrum for energy larger than 7 MeV. The multiplicity was compared with the results of Geant4 simulations. The Doppler corrected spectra were obtained for the 15.1 MeV gamma line, using the PSA+1HitID procedure.

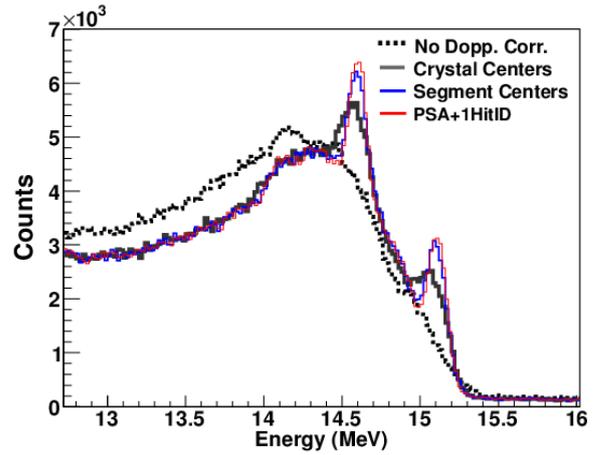

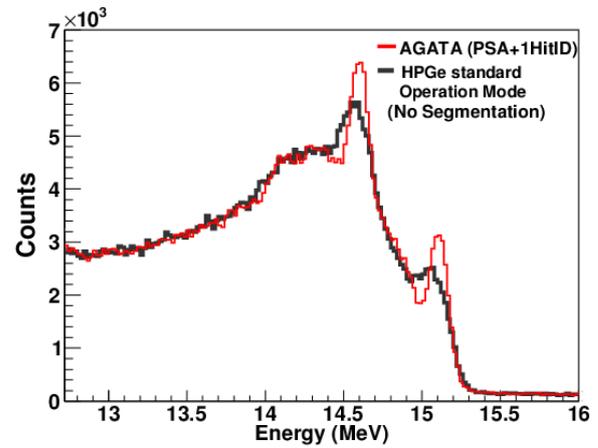

Figure 10. Gamma-ray spectra acquired during the in-beam test, displayed in the region around 15 MeV. In the top panel the spectrum without Doppler correction (dashed black line) is compared to: i) the spectrum obtained by applying a Doppler correction using only the central position of the HPGe crystal with the largest energy deposit (thick gray line), ii) the spectrum obtained using only the central position of segments (thin blue line) and iii) the spectrum obtained using the PSA+1HitID (thin red line). In the bottom panel only the spectra showing the performance of the detectors when operated in standard mode (Doppler correction using only the central position of the HPGe crystal) and using the PSA+1HitID are displayed.

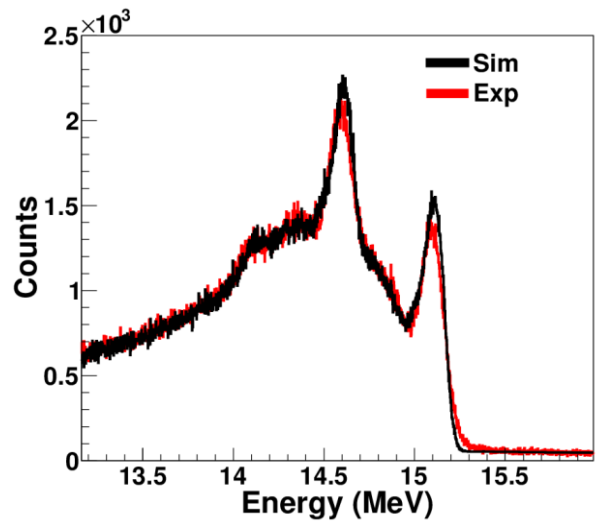

Figure 11. Comparison between experimental (red line) and simulated (black line) spectra in the 15 MeV region. The main factor limiting the FWHM of the 15.1 MeV gamma line was found to be the uncertainty due to the missing event by event reconstruction of the $^{12}$C ion velocity vector.

The main factor limiting the FWHM of the 15.1 MeV gamma line was found to be the uncertainty due to the missing event by event reconstruction of the $^{12}$C ion velocity vector. An intrinsic resolution of the order of 10 keV should be expected at the energy of 15 MeV. The simple add-back and PSA+1HitID algorithm, in the case of the 15.1 MeV gamma-rays, resulted to provide 4 times more counts in the full energy peak than the standard tracking algorithm. This is due to the fact that the 15.1 MeV gamma-ray has multiplicity 1, the level of background is low and that the tracking algorithm was optimized in the energy range 0-4 MeV where Compton scattering dominates; at 15 MeV the pair production is the main interaction mechanism instead. As in several AGATA physics cases which involve the measurement of high energy gamma-rays the "multiplicity =1" condition is fulfilled, the presented results might suggest a simple and efficient alternative to standard tracking, provided that the level of background radiation is sufficiently low.

**Acknowledgements**

This research has received funding from the European Union Seventh Framework Program FP7/2007-2013 under grant Agreement n° 262010 – ENSAR.
A. G. activity has been supported by the MINECO Spain, under grants AIC-D-2011-0746, FPA2011-29854 and by and Generalitat Valenciana, Spain, under grant PROMETEO/2010/101
We acknowledge the support by the German BMBF under Grants 06K-167 and 06KY205I.